\documentstyle[preprint,aps,epsf]{revtex}
\tighten

\begin{document}

\draft

\title{
Theoretical predictions for extraction of $\bbox{G_E^n}$ from 
semi-inclusive electron scattering on polarized $\bbox{^3}$He
based on various nucleon-nucleon interactions
}

\author{
J.~Golak$^{1,2}$,
W.~Gl\"ockle$^1$,
H.~Kamada$^3$
H.~Wita\l{}a$^2$, 
R.~Skibi\'nski$^2$,
A.~Nogga$^4$
}
\address{$^1$Institut f\"ur Theoretische Physik II,
         Ruhr Universit\"at Bochum, D-44780 Bochum, Germany}
\address{$^2$M. Smoluchowski Institute of Physics, Jagiellonian University,
                    PL-30059 Krak\'ow, Poland}
\address{$^3$ Department of Physics, Faculty of Engineering,
   Kyushu Institute of Technology,
   1-1 Sensuicho, Tobata, Kitakyushu 804-8550, Japan}
\address{$^4$ Department of Physics, University of Arizona, Tucson, 
              Arizona 85721, USA}
\date{\today}
\maketitle

\begin{abstract}
The process
$\overrightarrow{^3{\rm He}}(\vec{e},e'n)$
is theoretically analyzed with
the aim to search for sensitivity to the electric form factor of the
neutron, $G_E^n$. Faddeev calculations based on five high precision
nucleon-nucleon force models are employed and stability versus exchange 
of the nucleon-nucleon forces is demonstrated.
\end{abstract}

\narrowtext

In a recent paper~\cite{golak.02} we investigated the sensitivity of the process
$\overrightarrow{^3{\rm He}}(\vec{e},e'n)$ to the
extraction of the electric form factor of the neutron. 
That study was based on the AV18~\cite{AV18} nucleon-nucleon (NN) force 
and related MEC's and Faddeev calculations 
for $^3$He consistent with the final state continuum. 
We found that final state interaction (FSI) effects were quite 
important and an analysis without FSI cannot 
be recommended. Though we know from most of our experience that the theoretical 
results are quite
stable under exchange of one of the high precision NN potentials by another one,
we would like to supplement our previous paper~\cite{golak.02} 
by explicitely demonstrating
the independence from the NN interaction
for that specific process. We refer to~\cite{golak.02} for all
information about the formalism and the
definitions of the asymmetries $A_\perp$ and $A _\parallel$. 
Since in contrast to AV18 there are no consistent MEC's worked out for the other 
four high precision NN potentials CD Bonn~\cite{CDBonn},
Nijmegen 93, Nijmegen I and II~\cite{Nijm} we show only the asymmetries evaluated with
those NN potentials
based on the standard nonrelativistic single nucleon current operator. 

The important
ratio
$ A_\perp / A_\parallel $
for the
extraction of $G_E^n$ is displayed in Fig.~1 for the same $q^2$-values as in~\cite{golak.02}. 
This figure
corresponds to Fig.~8 of Ref.~[1]. It shows the five predictions (including AV18) 
for a fixed choice of $G_E^n$ as used in~\cite{golak.02}. In order to provide 
the necessary information on the sensitivity with respect to $G_E^n$ 
we include also three additional curves, where $G_E^n$ is
multiplied by 0.75, 1.25 and 1, respectively and this for the choice of AV18+MEC.
These additional three curves are the same as in Fig.~8 of Ref.~[1]. 
We see first of all in all cases a rather narrow spread of the five NN force predictions 
which document the expected stability
of the results  under exchange of one NN force by another one. However, we had to reduce
somewhat the energy range near the upper end of the neutron energies in order to keep that
spread small. Especially for the higher $q^2$-values the spread increases quite a bit for
the lower neutron energies (not shown). 
Experimentally it should be possible to concentrate on that
restricted upper energy range, where the cross section is anyhow largest.

Now narrow is
meant in relation to the variation of the predictions by changing $G_E^n$ by $\pm$ 25 \%.
From the three added curves 
we see that the shifts 
caused by those changes of $G_E^n$ is by far larger than the spread induced by varying
the NN forces. For example at $q^2$= 0.45 ${\rm (GeV/c)}^2$ and $E_n$= 260 MeV
the spread due to different NN potentials is about 13~{\%}, whereas the shifts 
caused by the $G_E^n$ variations reach 70~{\%}.
We did not repeat the calculations varying $G_E^n$ for the NN force predictions
with a single nucleon current alone since one can safely expect that corresponding
shifts would result as for AV18+MEC. This Fig.~1 shows that the sensitivity to a $G_E^n$
extraction would not suffer from a theoretical uncertainty induced by the choice of the
NN force. However, one also sees that MEC effects can only be neglected near the very upper
end of that neutron energy spectrum. They are quite significant for  $q^2$= 0.05 
- 0.2 ${\rm (GeV/c)}^2$. The filled square is the result for 
the scattering on a free neutron at rest. This is treated fully relativistically 
and will be referred to as the pure
neutron result.
Clearly an analysis of data with such an oversimplified picture would be meaningless.

For the sake of completeness we also display $A_\parallel $ 
and $ A_\perp$ separately in Figs.~2 and 3 
(corresponding to Figs.~5 and~6 of Ref.~[1]), 
now again, as in Fig.~1, restricted to that upper $E_n$ energy range.
The spread among all curves for $A_\parallel$ is quite small and MEC effects for AV18
are also minor. The filled square is the pure neutron result.
For $A_\perp$, which carries the dependence on $G_E^n$, we see again only small spreads.
MEC effects are quite noticeable
only for $q^2$= 0.1 - 0.2 ${\rm (GeV/c)}^2$. 
Clearly FSI is mandatory: the pure neutron result, which is reached by PWIAS 
calculations~[1] is far off. For the highest $E_n$-values $A_\parallel$ is
very stable under exchange of the NN forces (effects below 1~\%). This observable
is also insensitive to the $G_E^n$ variations (changes are below 2~\%).
That is why the ratio $ A_\perp / A_\parallel$ reflects the sensitivity 
of $ A_\perp$ to both effects.

Summarizing we conclude that theoretical uncertainties arising from replacing one of the
modern high precision NN potentials by another one in the ratio 
$ A_\perp / A_\parallel $ are much smaller than
the sought changes due to $G_E^n$ variations. 
MEC effects as evaluated in conjunction with AV18
decrease strongly at the upper end of the neutron spectra.
Measurements concentrated to the upper end of the neutron energies would be ideal
and could provide important information on $G_E^n$.

\acknowledgements
This work was supported by
the Deutsche Forschungsgemeinschaft (J.G.),
the Polish Committee for Scientific Research
under Grants No. 2P03B02818
and 2P03B05622,
and by NFS grant \# PHY0070858.
R.S. is a holder of scholarship from the Foundation for Polish Science
and acknowledges its financial support.
The numerical calculations have been performed
on the Cray T90 of the NIC in J\"ulich, Germany.


\begin{figure}[h!]
\leftline{\mbox{\epsfysize=180mm \epsffile{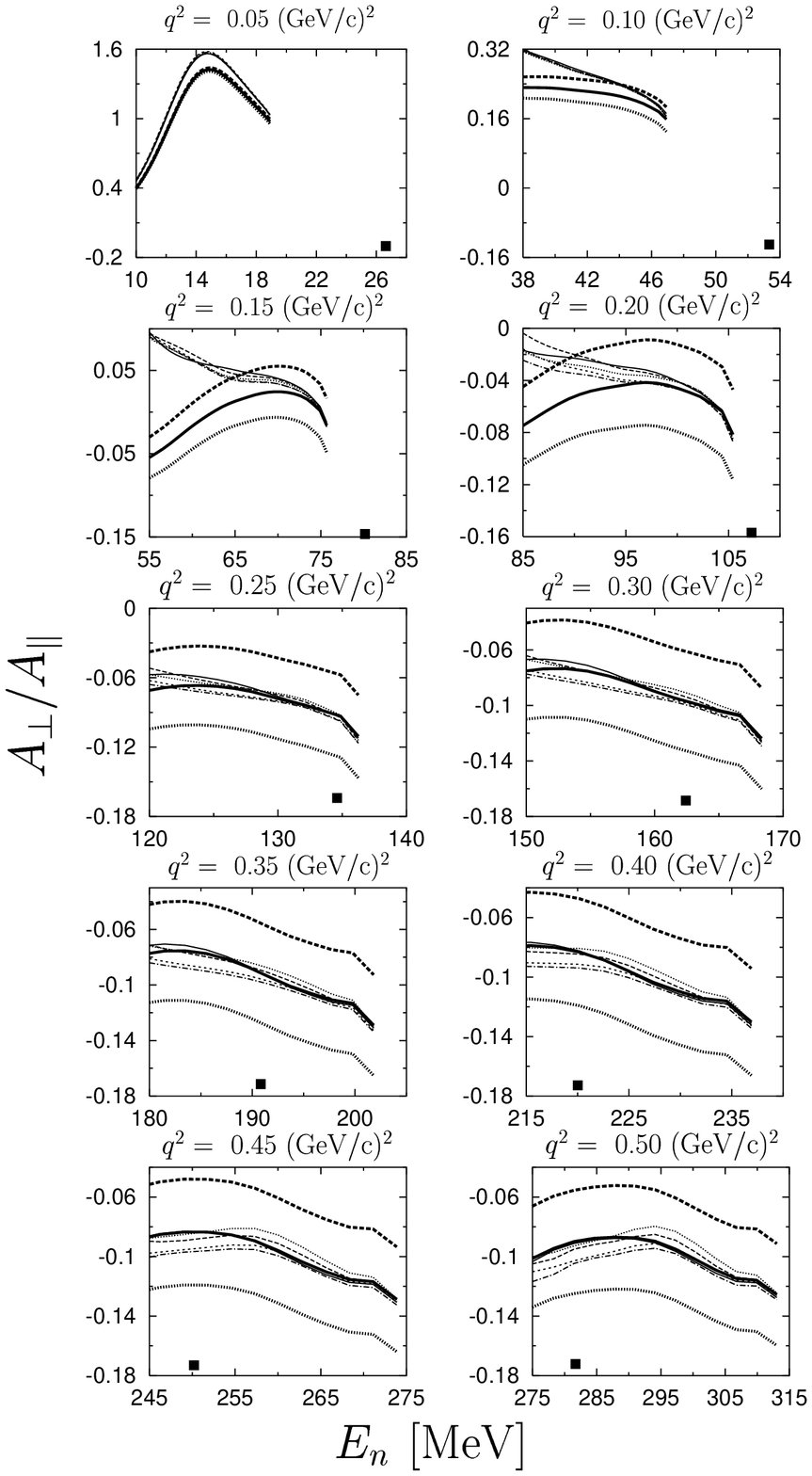}}}
\caption[ ]
{
The ratio $A_\perp / A_\parallel $
as a function of the neutron energy $E_n$
for different $q^2$-values. 
The thin lines correspond to full results (including FSI) without MEC's: 
CD Bonn (dash-dotted), Nijmegen~93 (dotted), Nijmegen~I (short dashed),
Nijmegen~II (long dashed) and AV18 (solid).
The thick lines are the full AV18 results including MEC's:
with 1.0$\, G_E^n$ (solid), with 0.75$\, G_E^n$ (dashed)
and with 1.25$\, G_E^n$ (dotted). 
The filled square is the pure neutron result.
}
\label{fig1}
\end{figure}

\begin{figure}[h!]
\leftline{\mbox{\epsfysize=180mm \epsffile{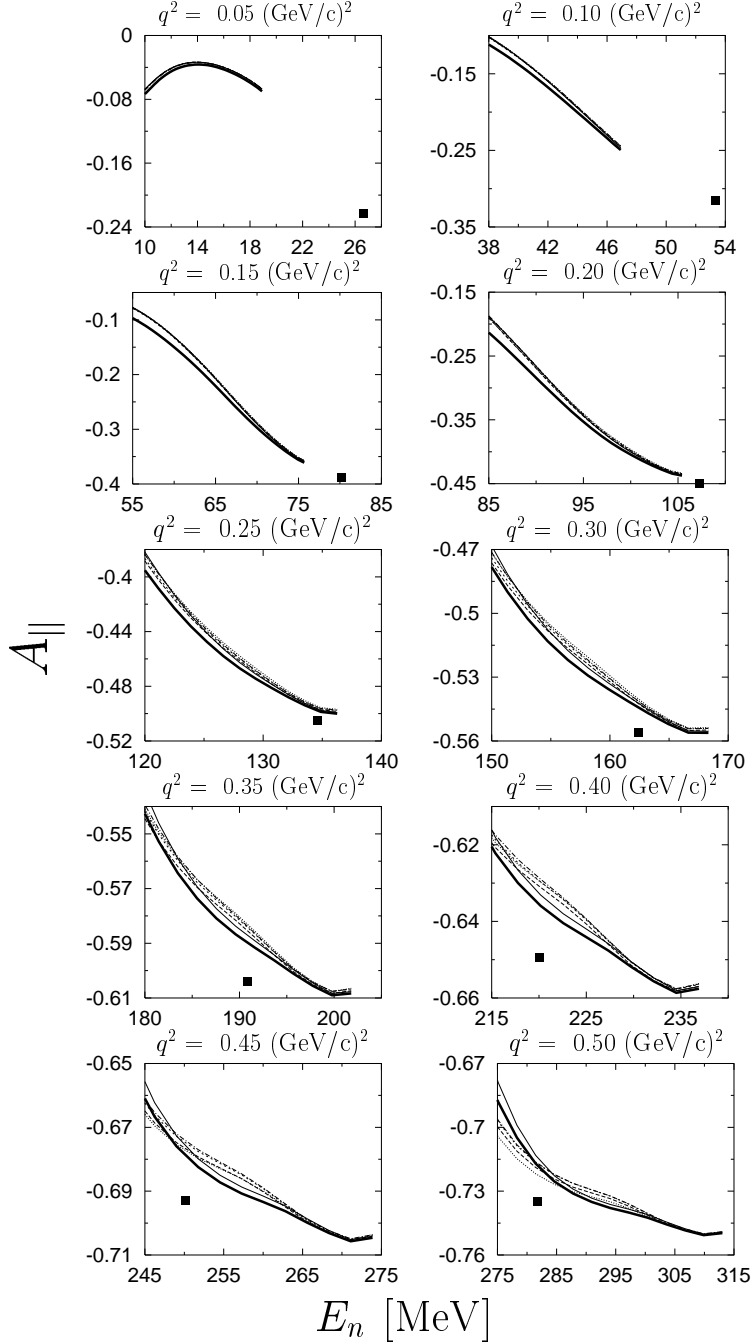}}}
\caption[ ]
{
$ A_\parallel $
as a function of the neutron energy $E_n$
for different $q^2$-values.
The thin lines and the symbol as in Fig.~1.
The solid thick line is the full AV18 result including MEC's.
}
\label{fig2}
\end{figure}

\begin{figure}[h!]
\leftline{\mbox{\epsfysize=180mm \epsffile{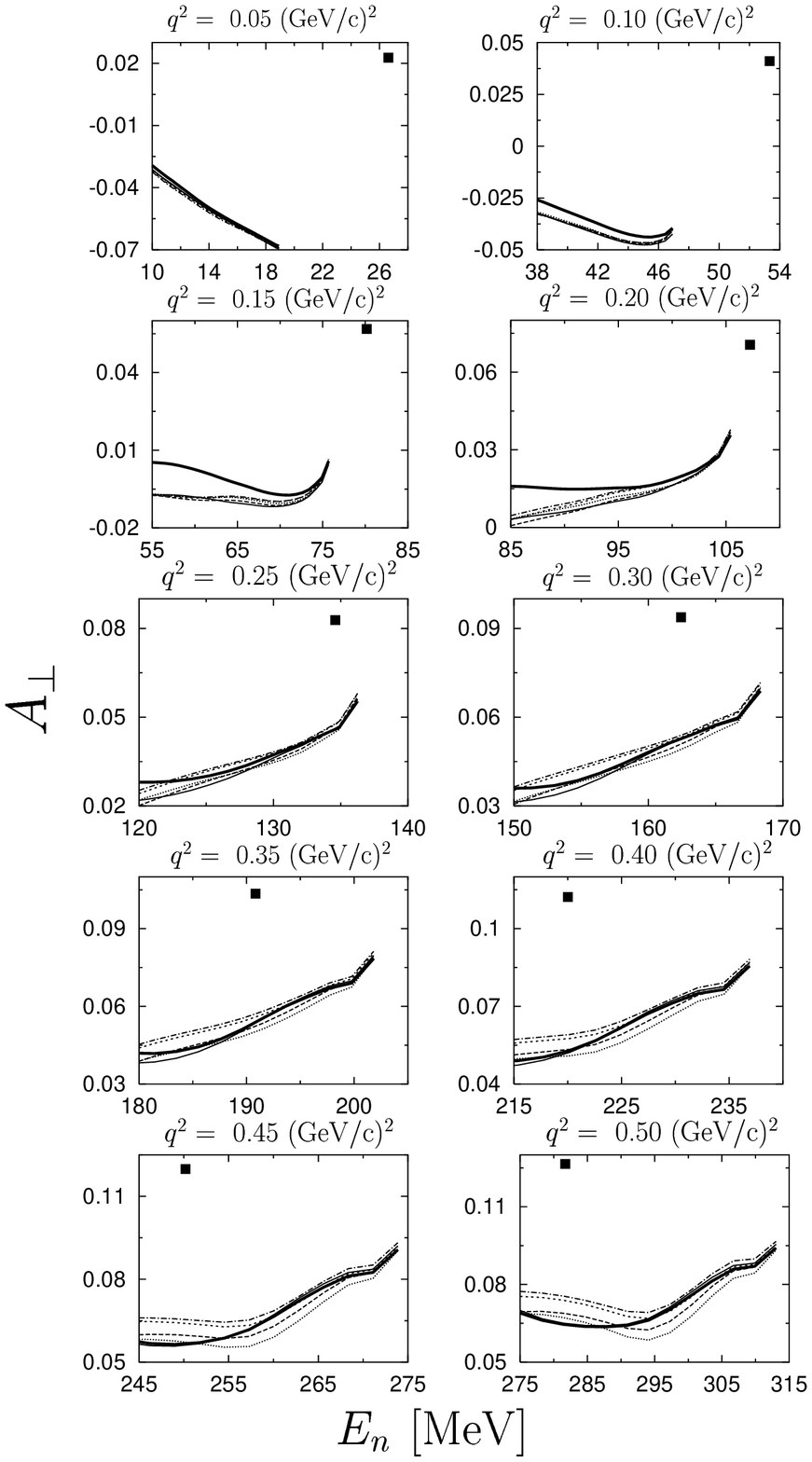}}}
\caption[ ]
{
$ A_\perp $
as a function of the neutron energy $E_n$
for different $q^2$-values.
Curves and the symbol as in Fig.~2.
}
\label{fig3}
\end{figure}


\begin{references}
\bibitem{golak.02} J.~Golak, W.~Gl\"ockle, H.~Kamada, H.~Wita{\l}a, 
                   R.~Skibi\'nski, and A.~Nogga,
                   Phys. Rev. C{\bf 65}, 044002 (2002).
\bibitem{AV18} R.~B.~Wiringa, V.~G.~J.~Stoks, R.~Schiavilla,
               Phys.~Rev.~C{\bf 51}, 38 (1995).
\bibitem{CDBonn} R.~Machleidt, F. Sammarruca, and Y. Song,
                 Phys. Rev. C{\bf 53}, R1483 (1996).
\bibitem{Nijm} V.~G.~J.~Stoks, R.~A.~M.~Klomp, C.~P.~F.~Terheggen,
               J.~J.~de Swart, Phys.~Rev.~C{\bf 49}, 2950 (1994).

\end{references}
\end{document}